\def\simgt{\lower 2pt \hbox{$\, \buildrel {\scriptstyle >}\over {\scriptstyle
\sim}\,$}}

\documentstyle[11pt,twoside,paspconf,epsfig]{article}
\renewcommand{\thepage}{\arabic{page}}
\begin{document}


\title{Ultrasoft Narrow-Line Seyfert~1 Galaxies: What Physical
Parameter Ultimately Drives the Structure and Kinematics of 
Their Broad Line Regions?}

\author{W. N. Brandt}
\affil{Department of Astronomy, The Pennsylvania State University, 
University Park, PA 16802}

\author{Th. Boller}
\affil{Max-Planck-Institut f\"ur Extraterrestrische
Physik, 85740 Garching, Germany}






\begin{abstract}
Recent X-ray observations have greatly advanced the  
study of ultrasoft Narrow-Line Seyfert~1 galaxies and have 
revealed strikingly clear correlations between optical emission 
line properties and the shape of the X-ray continuum. 
In particular, the strength of the soft X-ray excess relative 
to the hard X-ray power law appears to 
be directly related to the `primary eigenvector' of 
Boroson \& Green. Ultrasoft Narrow-Line Seyfert~1s
lie toward one extreme of the primary eigenvector, 
and their extreme X-ray spectral, X-ray variability and optical emission 
line properties suggest that they have extremal values of a primary physical 
parameter. We discuss efforts to identify this parameter and the 
observational evidence that it is the fraction of the Eddington rate 
at which the supermassive black hole is accreting. 
\end{abstract}




\section{Narrow-Line Seyfert~1s: An Extreme of Seyfert Activity}

\begin{figure}[t!]
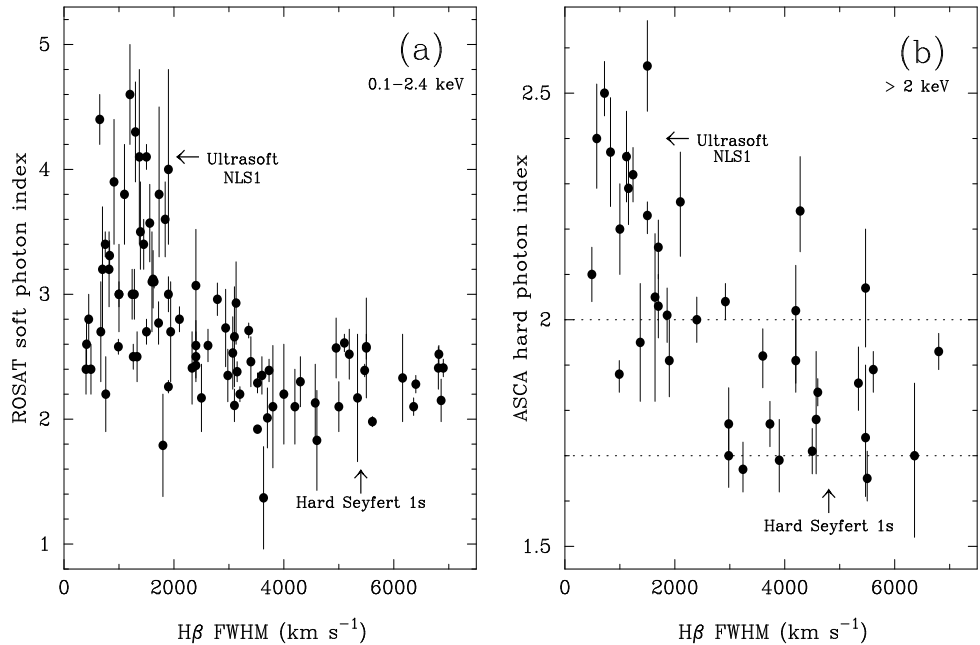

\hbox{
\epsfig{figure=brandt_fig1a.ps,height=8.5cm,width=6.2cm,angle=0}
\hspace{0.2 cm}
\epsfig{figure=brandt_fig1b.ps,height=8.5cm,width=6.2cm,angle=0}
}
\caption{Relations between X-ray continuum shape and H$\beta$ FWHM for
Seyfert~1 galaxies and QSOs. 
{\bf (a)} {\it ROSAT\/} photon index from a power-law fit 
versus H$\beta$ FWHM. In this 
band (0.1--2.4~keV), both the soft X-ray excess and the underlying 
power law contribute to the spectrum (compare with Fig.~3). The 
photon index serves to quantify the relative strengths 
of these two components. 
Note the generally steep NLS1 spectra and the apparent absence of 
objects with both steep soft X-ray slopes and H$\beta$ FWHM larger 
than $\approx 2000$~km~s$^{-1}$. 
{\bf (b)} {\it ASCA\/} hard X-ray photon index versus H$\beta$ FWHM (note 
that the ordinate scale is smaller than in the left hand panel). This
diagram has been constructed using only data above 2~keV, and hence
only the underlying power law contributes to 
the spectrum (again compare with Fig.~3). The horizontal dotted lines
show the `canonical' range of Seyfert~1 photon indices. Note that many
ultrasoft NLS1 have hard photon indices that lie above the `canonical'
range of 1.7--2.0. This shows that there is substantially more 
2--10~keV continuum diversity than was previously recognized. 
}
\end{figure}
 
Many Seyfert galaxies and QSOs show highly-luminous
and variable soft X-ray excess emission from their black hole regions. 
There is moderately good evidence that this emission is the high-energy
tail of the `Big Blue Bump' (BBB) 
that rises through the optical and ultraviolet. 
Together, the BBB and soft excess dominate the bolometric
luminosity and provide the primary source of energy for the 
strong, broad emission lines that were the focus of this
conference. 

Clear relations are now apparent between the strength of the
soft X-ray excess (relative to the underlying X-ray power law) and the
FWHM of the optical Balmer lines, particularly H$\beta$ 
(e.g., Boller, Brandt \& Fink 1996; Laor et~al. 1997).
This is shown in Fig.~1a, where the photon index from a power-law
fit to {\it ROSAT\/} data is used as a first-order way to quantify the
relative strengths of the soft excess and power law.  
All of the Seyferts/QSOs with the strongest and hottest soft X-ray
excesses are Narrow-Line Seyfert~1 class galaxies (NLS1).
NLS1 have relatively narrow Balmer lines with FWHM of only
500--2000~km~s$^{-1}$, and their interesting optical properties 
have been known for many years (e.g., Bergeron \& Kunth 1980; 
Gaskell 1985; Osterbrock \& Pogge 1985; 
Halpern \& Oke 1987). For example, NLS1 are prototypical
emitters of strong optical Fe~{\sc ii} emission, and they often 
have relatively weak [O~{\sc iii}] emission. 
Due to their strong soft X-ray emission, NLS1 are found in 
abundance in soft X-ray-selected surveys 
(e.g., Puchnarewicz et~al. 1992; Grupe et~al. 1998). 
They comprise a significant ($\sim$~20--30\%) but
often overlooked part of the Seyfert/QSO population. 

NLS1 do not appear to form a distinct class but are 
instead connected to the `standard' broad-line Seyfert~1s 
through a continuum of properties (e.g., Goodrich 1989).
In particular, NLS1 lie toward one extreme of the 
strong set of optical emission line correlations studied
by Boroson \& Green (1992a; hereafter BG92). These correlations 
emerged as the `primary eigenvector' of 
the Principal Component Analysis (PCA) 
performed by BG92, and they relate Balmer line
width/shape, Fe~{\sc ii} emission and [O~{\sc iii}] emission. 
The primary eigenvector represents the {\it strongest\/}
set of optical emission line correlations found among 
QSOs, and several ultraviolet line 
properties also appear to be connected with the primary eigenvector
(e.g., Wills et~al. 1998). 
BG92 suggested that their primary eigenvector was 
`driven' by an important, underlying physical parameter,
and they speculated that this parameter might be the fraction of 
the Eddington rate at which the supermassive black hole is 
accreting ($\dot M/\dot M_{\rm Edd}$). 

\begin{figure}[t!]
\hspace{0.8cm} \epsfig{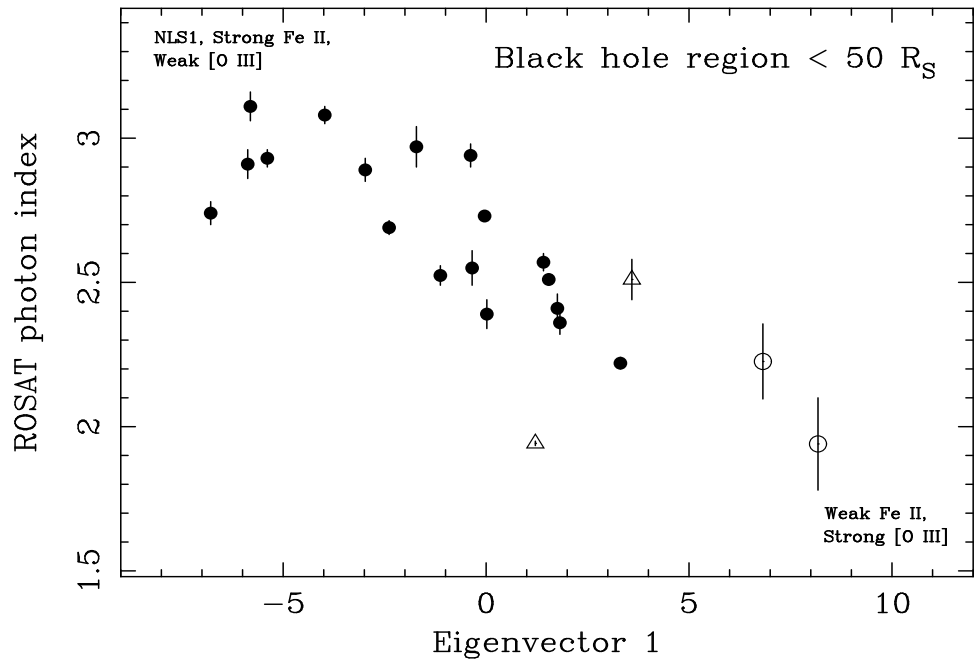}
\caption{{\it ROSAT\/}-band (0.1--2.4~keV) photon 
index versus the value of the BG92 
primary eigenvector for the PG QSOs from Laor et~al. (1997). This 
figure should be compared with fig.~2 of Boroson \& Green (1992b). 
Solid dots are radio-quiet QSOs, open triangles are core-dominated 
radio-loud QSOs, and open circles are 
lobe-dominated radio-loud QSOs. A Spearman rank-order 
test gives a Spearman $r_{\rm s}$ value
of $-0.834$, which is significant with over 99.9\%
confidence. This correlation is substantially stronger than
and probably responsible for the effect discussed by Corbin (1993).
The one outlying open triangle is 3C273, which 
appears to have a time-variable spectral shape and ionized absorption. 
}
\end{figure}

As expected from the discussion above, the {\it ROSAT\/}-band 
spectral slope also shows a remarkably strong 
correlation with the BG92 primary eigenvector (see Fig.~2). 
In fact, it appears that {\it ROSAT\/} spectral slope correlates
more tightly with the primary eigenvector than with any of the
individual optical emission line properties used to define
the eigenvector (see Brandt \& Boller 1998 for details). Since the 
highly-luminous and rapidly-variable X-ray emission is thought to be 
formed within $\sim 50$ Schwarzschild radii ($50R_{\rm S}$)
of the supermassive black hole, the correlation suggests that the 
physical parameter driving the eigenvector also ultimately 
originates within $\sim 50R_{\rm S}$. A compact origin is certainly
expected if $\dot M/\dot M_{\rm Edd}$ drives the eigenvector, 
but it is not proper proof of this
hypothesis. One would like to find compelling evidence that 
clinches the $\dot M/\dot M_{\rm Edd}$ interpretation. 
This would be an important advance because:

\vspace{-0.05truein}

\begin{itemize}

\vspace{-0.05truein}

\item We would gain a way to gauge the primary driver
of these accretion-powered sources. 

\vspace{-0.05truein}

\item We would determine the root cause of the extreme X-ray 
spectral and variability properties of ultrasoft NLS1. 

\vspace{-0.05truein}

\item We would isolate one of the main effects shaping the 
structure and kinematics of the Broad Line Region (BLR) and
Narrow Line Region (NLR). 

\end{itemize}

\vspace{-0.05truein}

\noindent
While compelling evidence has not yet been found, it is clear 
that research directed towards this goal is driving our theoretical 
understanding of Seyferts and QSOs more generally. In the rest of
this article, we will describe some of the progress that has been made 
and briefly examine future prospects. 


\section{Optical Emission Line Properties}

The most notable optical property of ultrasoft NLS1 is their narrow
Balmer lines, and these can be plausibly explained in terms of a model 
which postulates high $\dot M/\dot M_{\rm Edd}$
(e.g., Boller, Brandt \& Fink 1996; Laor et~al. 1997). 
The basic assumptions of this model are that
(1) the motions of the BLR clouds are virialized 
in the gravitational potential of the central black hole 
and 
(2) the size of the BLR is primarily a function of 
luminosity (perhaps due to the sublimation of dust;
cf. Netzer \& Laor 1993). 
Consider a set of QSOs that have similar luminosities and
hence have BLRs of about the same size. The objects of this set
with higher $L/L_{\rm Edd}$ will have smaller-mass black holes.
Hence the virialized motions of their BLR 
clouds will have smaller velocity and their BLR lines will 
be narrower (see section~4.7 of Laor et~al. 1997
for the corresponding equations). 
 
While this basic picture admittedly lacks the details 
present in more sophisticated models of the BLR, it 
does demonstrate that there is a plausible connection between 
Balmer line width and $\dot M/\dot M_{\rm Edd}$. 
If one also takes into account the shape of the ionizing 
continuum, this further enlarges the relative 
sizes of the BLRs of ultrasoft NLS1 and strengthens the 
dependence on $\dot M/\dot M_{\rm Edd}$ 
(see Wandel 1997). 
The currently available observations of optical line variability 
appear to be consistent with the basic picture outlined above 
(see Giannuzzo et~al. 1998). 

It is more difficult at present to state a direct, simple reason 
why strong optical Fe~{\sc ii} emission and weak [O~{\sc iii}] 
emission should arise when $\dot M/\dot M_{\rm Edd}$ is large. 
However, some suggestions have been put forward 
(e.g., Boroson \& Green 1992b) and need further investigation. 
We briefly return to the topic of Fe~{\sc ii} below when describing 
the power-law X-ray continua of ultrasoft NLS1.


\section{Broad-Band X-ray Continua}

The X-ray luminosities of most ultrasoft NLS1 appear to be dominated 
by the emission from the strong and hot ($\simgt 100$~eV) soft X-ray 
excess. This component can dominate ultrasoft NLS1 X-ray 
spectra up to $\approx 1.5$~keV, and it can be acceptably 
modeled (at least in a statistical sense) using blackbodies or 
multi-temperature accretion disk models. The spectral character of this
radiation should reflect the structure, temperature and dynamics of the
inner accretion disk. However, current observational limitations have 
hindered the precision testing of physical models for the soft excess. 
These limitations should be removed over the next few years with the 
launches of the next generation of X-ray observatories. 
 
\begin{figure}[t!]
\hspace{0.8cm} \epsfig{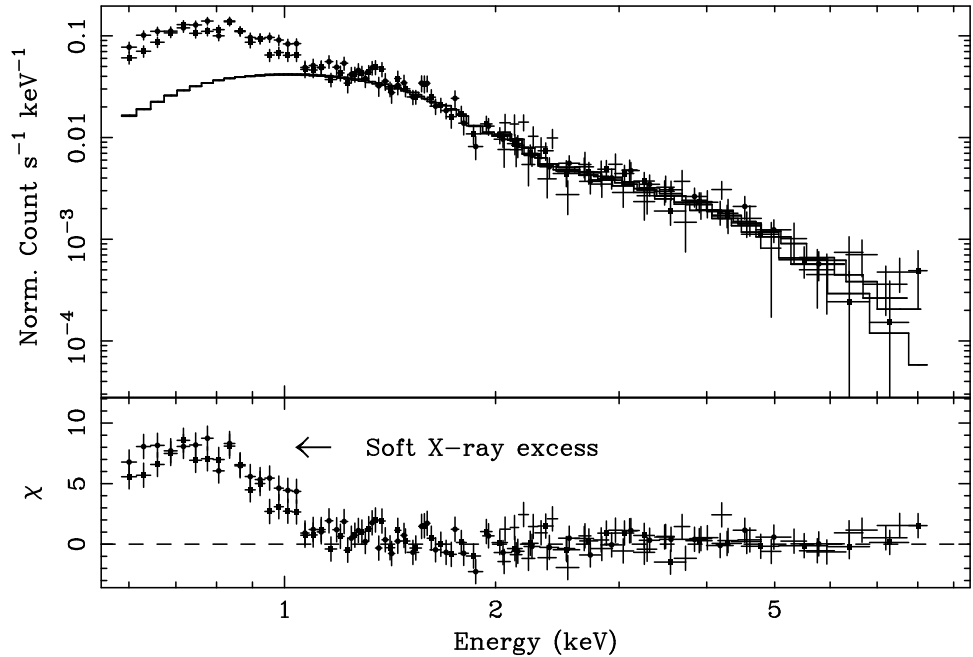}
\caption{{\it ASCA\/} spectra of the ultrasoft NLS1 H~0707--495. A power-law model has 
been fit to the data above 2~keV and extrapolated downward in energy to show 
the systematic deviations from the model at low energies. Note the prominent soft 
X-ray excess emission that dominates the spectrum below $\approx 1.1$~keV. This
emission is even stronger at energies below the {\it ASCA\/} band. The 
ordinate for the lower panel (labeled $\chi$) shows the fit residuals in 
terms of $\sigma$ with error bars of size one.}
\end{figure}
 
The $>2$~keV spectra of ultrasoft NLS1 were poorly studied until
recently, largely due to the fact that few of these objects were detected
in hard X-ray sky surveys. {\it ASCA\/} and {\it SAX\/} have allowed the
first systematic studies of ultrasoft NLS1 in the 2--10~keV band. 
While their spectra do appear to `break' to power laws above 2~keV
(see Fig.~3, for example), these power laws show a wider range of slopes 
than has been seen among hard X-ray-selected type~1 objects 
(Brandt, Mathur \& Elvis 1997; see Fig.~1b). Steep power-law photon 
indices of $\approx$~2.0--2.6 appear to be fairly common for ultrasoft 
NLS1, and these directly demonstrate that there is substantially more 
2--10~keV continuum diversity among Seyferts and QSOs than was 
previously recognized. The origin of this spectral diversity needs 
to be better understood. Since the hard X-ray power law is thought to 
be produced by the Compton upscattering of disk photons in an accretion 
disk corona, its photon index should provide information about the 
Thomson depth, temperature, and bulk motions of the corona. A steep 
2--10~keV power law could arise as the result of 
a thinner corona (e.g., Poutanen, Krolik \& Ryde 1997),
a cooler corona (e.g., Pounds, Done \& Osborne 1995), or
a corona with large bulk motions (e.g., Ebisawa, Titarchuk \& Chakrabarti 1996). 

The strong soft X-ray components and steep power laws seen from many 
ultrasoft NLS1 are reminiscent of Galactic black hole candidates (GBHC) 
accreting in their ultrasoft high states. When GBHC make transitions to 
their high states, their hard photon indices increase from
1.5--2.0 to 2.0--2.5 (compare with Fig.~1b). If an analogy between 
high state GBHC and ultrasoft NLS1 is 
appropriate, then the steep power laws of 
ultrasoft NLS1 again suggest that these systems are accreting with 
relatively high $\dot M/\dot M_{\rm Edd}$. 

The steep $>2$~keV continua of ultrasoft NLS1 with strong 
optical Fe~{\sc ii} emission also have implications 
for models of radiative Fe~{\sc ii} formation. 
In the absence of any strong spectral flattening above the 
{\it ASCA\/} band, the total hard X-ray emission from these objects 
will be weak. This provides evidence against models of radiative 
Fe~{\sc ii} formation that require strong hard 
X-ray emission such as the Compton-heating model of 
Collin-Souffrin, Hameury \& Joly (1988). Other radiative 
heating models predict an anticorrelation between 
Fe~{\sc ii}/H$\beta$ and hard X-ray luminosity 
(see section~4 of Joly 1993 and references therein), and 
this appears to be what the currently available 
X-ray data suggest. 


\section{X-ray Variability}

Strong variability in an ultrasoft NLS1 appears to have been first 
noted by Zwicky (1971), who commented on optical `outbursts' in 
I~Zwicky~1 due to `implosive and explosive events.' He urged 
monitoring of this variability, and he would perhaps be happy 
to see that X-ray monitoring of ultrasoft NLS1 (also sometimes referred 
to as `I~Zwicky~1 objects') has recently proven quite fruitful. 

\begin{figure}[t!]
\hspace{0.8cm} \epsfig{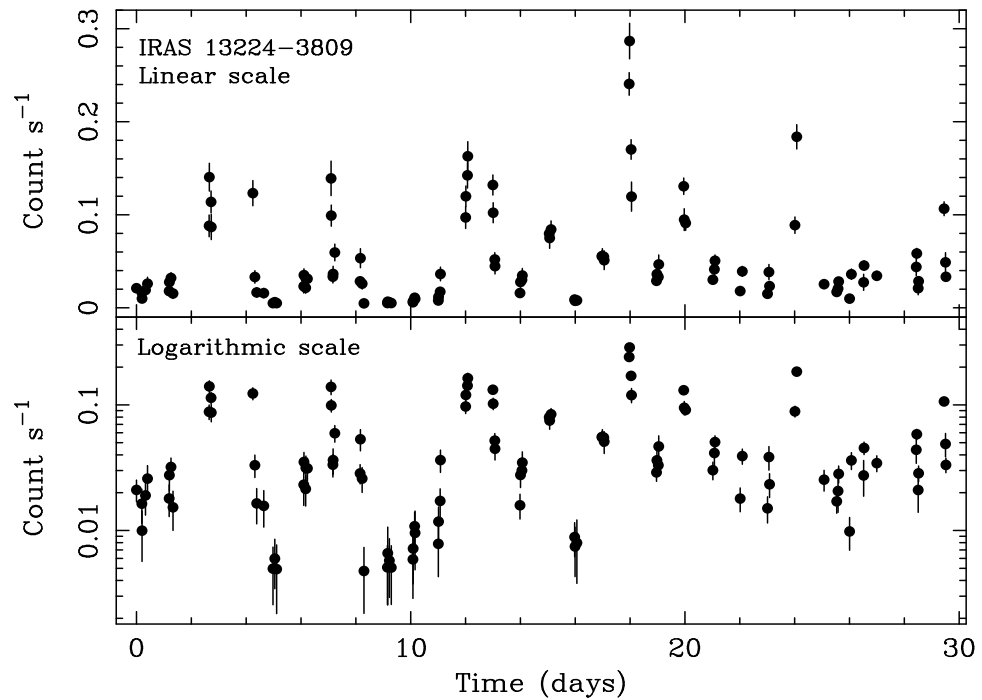}
\caption{{\it ROSAT\/} HRI light curve for IRAS~13224--3809 obtained during a 
30-day monitoring campaign in 1996~January--February. The light curve 
is shown with both linear and logarithmic scalings of the 
ordinate. By holding this book upside down (and perhaps using a mirror), you 
can judge if the `logarithmic' light curve 
looks symmetric (see the question by Martin Gaskell in the discussion section).}
\end{figure}

Ultrasoft NLS1 generally (but not always) appear to 
exhibit the most extreme X-ray variability found 
among radio-quiet Seyferts/QSOs. For example, Boller et~al. (1997) 
have discovered persistent, giant-amplitude (factors of 35--60) 
X-ray variability events from the 
ultrasoft NLS1 IRAS~13224--3809 (see Fig.~4; also see
Otani, Kii \& Miya 1996). The observed light curve consists of 
multiple strong flares interspersed among relatively quiescent 
periods, a characteristic of nonlinear variability (see Vio et~al. 1992), 
and Boller et~al. (1997) have demonstrated that the variability is in fact 
nonlinear.\footnote[1]{Nonlinear X-ray variability has also been seen 
in other ultrasoft NLS1 including 
PKS~0558--504 (Remillard et~al. 1991) and 
NGC~4051 (Green, McHardy \& Done 1998).} 
This is physically important because it rules out the possibility 
that the variability is produced by the simple linear superposition
of a large number of independent emission regions; rather the emission 
regions must interact in a nonlinear manner. 
Another impressive example of extreme X-ray variability has recently 
been provided by an 18-day HRI monitoring campaign on the highly-luminous 
NLS1-class QSO PHL~1092 (Brandt et~al., in preparation). Based on 
reports of strong X-ray variability in a short {\it ROSAT\/} PSPC
observation (e.g., Forster \& Halpern 1996), we monitored PHL~1092
and found remarkably rapid and large-amplitude X-ray variability 
for such a luminous object (see Boller \& Brandt 1998 for details). 
When `efficiency limit' arguments (see Fabian 1979) are 
applied to the most rapid observed variability 
event, relativistic X-ray flux boosting appears to be implied. 
Finally, systematic studies of Seyfert/QSO variability 
(e.g., Green, McHardy \& Lehto 1993; Fiore et~al. 1998a) suggest that
ultrasoft NLS1 as a class show enhanced X-ray variability
over a wide range of timescales. 

The cause of the extreme X-ray variability of ultrasoft NLS1 is not
properly understood. Enhanced X-ray variability might again plausibly 
arise as the result of high $\dot M/\dot M_{\rm Edd}$,\footnote[2]{At a 
fixed luminosity, black holes radiating at  
higher fractions of the Eddington rate will have lower masses. 
Lower mass black holes are thought to be associated with physically 
smaller emission regions that vary more rapidly.}
but in this case there does not appear to be a direct, simple analogy
with GBHC in their ultrasoft high states (see the response to 
Mike Eracleous in the discussion section). In addition, it 
is not obvious that a high $\dot M/\dot M_{\rm Edd}$ would lead to 
a predilection for nonlinear flaring or the relativistic boosting 
of X-rays. There is an eerie similarity between the nonlinear flaring
seen from ultrasoft NLS1 and that seen from many blazars, and some
have even suggested that ultrasoft NLS1 may harbor small relativistic 
X-ray jets that intermittently undergo nonlinear flares
(Green, McHardy \& Done 1998). It is perhaps not a coincidence
that sources with steep energy spectra are those most prone 
to relativistic variability enhancement when observed in a fixed 
energy band (see Boller et~al. 1997 for a detailed discussion). 


\section{Spectral Features}

The 6.40--6.97~keV iron~K$\alpha$ lines observed from type~1 Seyferts/QSOs 
are often associated with the innermost parts of accretion disks. Theoretical 
calculations generally predict that an accretion disk will become more highly 
ionized as $\dot M/\dot M_{\rm Edd}$ increases (e.g. Ross \& Fabian 1993), and 
the ionization of iron should be detectable as 
an increase in the centroid energy of the iron~K$\alpha$ line. 
Unfortunately, current line measurements do not show a clear trend 
towards increasing iron ionization for ultrasoft NLS1. Some 
ultrasoft NLS1 do show evidence for ionized iron~K$\alpha$ lines, but others do not
(e.g., Comastri et~al. 1998). It is important to note, however, that iron 
line fitting is currently very difficult for ultrasoft NLS1. The photon
statistics for the observed lines are limited, and as a result line fits 
can be degenerate and difficult to interpret. A more unified picture may 
well emerge based on observations with satellites such as 
{\it XMM\/}, {\it Astro-E\/} and {\it Constellation-X\/}.  

Even less clear at present is the origin of the peculiar spectral 
residuals seen from 1.0--1.2~keV in some ultrasoft NLS1 
(e.g., Brandt et~al. 1994; Leighly et~al. 1997; 
Fiore et~al. 1998b; Turner, George \& Nandra 1998). These 
can be statistically modeled using edges 
and/or lines, but standard ionized X-ray 
absorber models cannot produce edges/lines 
at the correct energies without also producing much stronger features at
other energies which are not observed. Brandt et~al. (1994) pointed
out, with due caution, that the residuals could in principle be 
explained by Doppler blueshifted absorption from ionized oxygen
(with a velocity of $\sim 0.3c$), and this possibility was 
explored in detail by Leighly et~al. (1997) using {\it ASCA\/} 
data. Detailed modeling is difficult at present since the 
soft X-ray excess and hard power law are exchanging dominance of the 
spectrum near the energy where the putative features are observed. 
In addition, the sources with the residuals often show significant 
spectral variability and this can lead to confusion when spectra are
binned using time intervals that include such variability.
Another possibility is that at least some of these features arise
from reflection off a highly ionized inner accretion disk, and as 
mentioned above such a disk is expected for high 
$\dot M/\dot M_{\rm Edd}$. These features definitely need 
further study; they may well offer a new probe of the nuclear environment 
that has yet to be properly exploited. 


\section{Other Possible Drivers of the Primary Eigenvector}

An alternative way to clinch the high $\dot M/\dot M_{\rm Edd}$ interpretation
of the primary eigenvector would be to falsify all other plausible
interpretations. Alternatively, one of the other interpretations might
be proven correct! The two leading contenders at present are that
the primary eigenvector is driven by (1) orientation of the nuclear 
region and (2) black hole spin. 

BG92 presented a strong argument that orientation of the nuclear 
region could not drive the primary eigenvector. 
However, their argument relies upon 
the assumption that [O~{\sc iii}] emission is 
isotropic. This common assumption has recently been questioned
(e.g., Hes, Barthel \& Fosbury 1993; Baker 1997). Precise measurements of 
[O~{\sc ii}] emission, which has a lower critical density and ionization
potential than [O~{\sc iii}], can better test whether the primary eigenvector 
might be driven by orientation. Some of the requisite observations have been 
made, and analysis will be performed soon. 

There do not appear to be any compelling arguments against 
the possibility that black hole spin drives the primary 
eigenvector. BG92 noted that radio-loud QSOs fall towards
one end of their primary eigenvector, and based on this they 
suggested that the parameter driving the eigenvector was largely
responsible for the presence of radio emission
(see fig.~2 of Boroson \& Green 1992b). Black hole
spin has also been discussed in connection with the radio-loudness
`volume control' (e.g., Wilson \& Colbert 1995), with radio-loud
QSOs being postulated to have 
the most rapidly spinning black holes. Since
radio-loud QSOs and ultrasoft NLS1 lie at opposite ends of the
primary eigenvector, a spin interpretation for the eigenvector
would suggest that ultrasoft NLS1 have the slowest spinning 
black holes. However, it does not seem 
that a slowly spinning black hole
would naturally explain the observed properties of ultrasoft NLS1. 
If anything, their hot soft X-ray components and extreme X-ray
variability might suggest a rapidly spinning black hole
(e.g., Cunningham 1975 and Boller et~al. 1997). Observations 
of relativistic iron~K$\alpha$ lines with {\it XMM\/}, {\it Astro-E\/} 
and {\it Constellation-X\/} are needed to measure 
black hole spins and address this issue. 


%

%
%


\acknowledgments

W.~N.~Brandt gratefully acknowledges support from 
NASA grants NAG5-4826 and NAG5-6023. We thank S.~Gallagher
for helpful discussions. 


%
%




\begin{question}{Mike Eracleous}
You drew an analogy between the X-ray spectra of ultrasoft
NLS1 and the X-ray spectra of Galactic black hole candidates 
(GBHC) in their ultrasoft high states. Can one draw a similar 
analogy between the variability properties of the two classes? 
\end{question}
\begin{answer}{Niel Brandt}
I am not aware of a direct, simple analogy 
and suspect that it will be difficult to 
construct one. The soft components of high-state GBHC
show markedly reduced variability from the usual millisecond 
flickering, although they do vary on timescales of about a day. 
One might expect the supermassive analogs of high-state GBHC
to be stable on timescales of thousands of years 
if the variability scales roughly with black hole mass. 
However, this is not observed to be the case; the soft components
of many ultrasoft NLS1 show large-amplitude X-ray variability
on timescales down to $\sim 1000$~s. So the timescale 
discrepancy here is a factor of $\sim 10^7$! 

Some GBHC do have `very high states' which show rapid variability. 
Ultrasoft NLS1 could perhaps be in states analogous to these, 
although this complication causes the analogy to lose some of its 
directness and simplicity. 
\end{answer}

\vspace{0.3 cm}


\begin{question}{Martin Gaskell}
If you take an optical light curve plotted in magnitudes (i.e. on a logarithmic 
scale), you can turn it upside down and it looks the same. Your X-ray light curve
for IRAS~13224--3809 was plotted with a linear scale. Would it be more 
symmetric if you plotted it with a logarithmic scale? 
\end{question}
\begin{answer}{Niel Brandt}
Please see Fig.~4 which can be used to judge this matter. 
\end{answer}

\vspace{0.3 cm}


\begin{question}{Greg Shields}
Is the soft X-ray excess significantly better fit by a blackbody than by
optically-thin thermal bremsstrahlung? How do you get a blackbody much 
hotter than the effective temperature? 
\end{question}
\begin{answer}{Niel Brandt}
For the ultrasoft NLS1 that I know best, a blackbody model does provide 
a statistically better fit than an optically-thin thermal bremsstrahlung 
model. Optically-thin thermal bremsstrahlung also appears unlikely based
on {\it EUVE\/} results for the NLS1 Mrk~478 (Marshall et~al. 1996). 
Finally, the arguments of Elvis et~al. (1991) combined with the very 
rapid observed variability provide additional evidence against 
optically-thin thermal bremsstrahlung models. It is important to 
comment, however, that observational limitations currently hinder 
testing of all but the simplest spectral models for the soft X-ray excess. 
In addition, some objects appear to have soft X-ray excesses that are too 
broad to be fit by a single blackbody. These can be fit with a sum of 
blackbodies or a multi-temperature accretion disk model. 

I do not have a compelling theoretical explanation for the high  
blackbody temperatures. All I can say is that, from an observational point
of view, they appear to be present. The spectrum shown in Fig.~3, for 
example, gives a blackbody temperature of $\simgt 100$~eV. Relativistic effects
and Compton scattering can raise the apparent temperature. However, it 
is not clear that these effects can produce the observed spectral form. 
Better soft excess spectra from missions such as {\it AXAF\/} and {\it XMM\/} 
are needed for precision model testing. 
\end{answer}

\vspace{0.3 cm}


\begin{question}{Amri Wandel}
In order to constrain disk-corona models one would want to compare the
relative variability of the soft and hard X-ray components. Have you
tried to do such a comparison? 
\end{question}
\begin{answer}{Niel Brandt}
Detailed comparisons of this type are difficult at present since most current 
satellites do not provide simultaneous, precision constraints on both 
the soft and hard X-ray emission. 
%
%
%
%
The limited data currently available suggest that the soft emission
generally shows larger-amplitude variability than the hard emission. 
\end{answer}


\end{document}